\documentclass[12pt]{article} %***
\usepackage{amsmath}
\usepackage{amssymb}
\usepackage{amsfonts}
\usepackage{multirow}
\usepackage{natbib}
\usepackage{amsthm}
\usepackage{graphicx}
\usepackage{booktabs}
\usepackage[tight]{subfigure}
\usepackage{caption}
\usepackage{url}
\usepackage{hyperref}
\usepackage{setspace}
\usepackage[svgnames,usenames,dvipsnames]{xcolor}
\usepackage{graphicx}
\usepackage[margin=1in]{geometry}
\usepackage{float}
\usepackage{listings}
\usepackage{rotating}
\usepackage{subfigure}
\usepackage{tikz}
\usepackage{bbm}

\newcommand{\be}{\begin{equation}}
\newcommand{\ee}{\end{equation}}
\newcommand{\bea}{\begin{eqnarray}}
\newcommand{\eea}{\end{eqnarray}}
\newcommand{\mbf}[1]{\mathbf{#1}}

\newcommand{\indicator}{\mathbf{1}}

\newcommand{\bbs}{\mbf{s}}

\newcommand{\ben}{\begin{equation*}}
\newcommand{\een}{\end{equation*}}
\newcommand{\bean}{\begin{eqnarray*}}
\newcommand{\eean}{\end{eqnarray*}}
\newcommand{\bsm}{\begin{smallmatrix}}
\newcommand{\esm}{\end{smallmatrix}}
\newcommand{\bmat}{\begin{matrix}}
\newcommand{\emat}{\end{matrix}}

\newcommand\indsim{\stackrel{ind}{\sim}}
\newcommand\iidsim{\stackrel{iid}{\sim}}

\usepackage[left]{lineno}
\usepackage{letltxmacro}
\LetLtxMacro{\originaleqref}{\eqref}
\renewcommand{\eqref}{Eq.~\originaleqref}

\begin{document}

   \begin{center}
       \vspace*{1cm}
       \large
	    \textbf{Within-fire estimates of biomass loss across national forests of the US West Coast: inventory-informed inference with uncertainty}\\
        \normalsize
          \vspace{5mm}
	    Elliot S. Shannon\textsuperscript{1,2}, Andrew O. Finley\textsuperscript{1,2,*}, Paul B. May\textsuperscript{3}, Hans-Erik Andersen\textsuperscript{4}, Harold Zald\textsuperscript{5}, Grant M. Domke\textsuperscript{6} and George C. Gaines III\textsuperscript{7}
	   \\
        \vspace{5mm}
   \end{center}
   {\small 
   \begin{enumerate}
       \item Department of Forestry, Michigan State University, East Lansing, MI, USA.
       \item Department of Statistics and Probability, Michigan State University, East Lansing, MI, USA.
       \item Department of Electrical Engineering and Computer Science, South Dakota School of Mines \& Technology, Rapid City, SD, USA.
       \item USDA Forest Service, Pacific Northwest Research Station, Winthrop, WA, USA.
       \item USDA Forest Service, Pacific Northwest Research Station, Corvallis, OR, USA.
        \item Northern Research Station, United States Department of Agriculture Forest Service, St. Paul, MN, USA.
        \item USDA Forest Service, Rocky Mountain Research Station, Missoula, MT, USA.
       \end{enumerate}
       }
       \noindent $^\ast$ Corresponding Author: Andrew O. Finley, email: finleya@msu.edu.

\section*{Abstract}
Wildfire is reshaping forests in the western United States (US), yet estimates of wildfire impacts often rely on burned-area summaries or satellite-derived severity metrics that do not directly quantify changes in forest biomass or their uncertainty. We combine US Forest Inventory and Analysis plot measurements with annual 30 m tree canopy cover data in a Bayesian spatio-temporal modeling framework to estimate within-fire changes in live aboveground biomass (AGB) and forest mortality area across National Forest System lands in Washington, Oregon, and California. The model represents both live forest presence or absence and AGB conditional on live forest, allowing posterior predictive estimates of pre- and post-fire conditions at the pixel scale. Applied to 6,239 fires from 2000 to 2022, the analysis estimates 95.3 million Mg of wildfire-associated AGB loss and 2.39 million ha of forest mortality area, with the largest aggregate impacts occurring in 2020. Pixel-level estimates reveal substantial spatial heterogeneity within fire perimeters, including large differences between total burned area, forest mortality area, and associated biomass loss. This inventory-informed, uncertainty-aware approach provides a scalable framework for quantifying forest carbon impacts of disturbance across broad regions while retaining within-fire spatial detail.

\section{Introduction}\label{sec:intro}

Wildfire is a major driver of forest change and disturbance on the West Coast of the United States (US) \citep{marlon_wildfire_history}, but quantifying its impacts on forest structure remains a significant statistical challenge. Existing approaches rely primarily on satellite-derived metrics, such as the Relative Differenced Normalized Burn Ratio (RdNBR) \citep{Miller_2009}, to characterize burn severity and extent. However, these metrics are indirect proxies for vegetation change, do not support statistically robust estimation of forest structural change with associated uncertainty quantification, and are not well suited for estimating derived quantities of interest \citep{fire_severity_improvements}. In contrast, nationwide forest inventory (NFI) data provide plot-based estimates of important forest structure attributes, including live forest aboveground biomass (AGB) \citep{westfall_2023}. Due to cost and logistical constraints, however, NFI data are spatially and temporally sparse, limiting reliable estimation of forest structure attributes except over large aggregated areas \citep{need_for_small_area_estimation}. Thus, remote sensing products alone are insufficient because they do not directly support inference on forest structural change, while NFI data alone are too sparse to support fine-scale assessment of individual wildfires. Quantifying wildfire impacts at fine spatial and temporal scales therefore requires statistical methods that integrate both sources of information within a unified inferential framework.

This need is especially important because forests provide a wide range of ecosystem services, including timber production, improved water and air quality, and wildlife habitat \citep{sanchez_ecosystem_services}. In the context of climate change, forests also function as a major carbon sink in the US, sequestering atmospheric carbon in live woody biomass through photosynthesis \citep{WOODBURY200714} and retaining it in standing and down dead wood \citep{wear2015sink}. Over the past several centuries, forest management has heavily emphasized fire suppression, increasing the potential for more severe wildfires \citep{low_intensity_fires_mitigate_high_intensity_fires, Kreider_2024}. Today, accumulated fuels together with more extreme climatic conditions continue to exacerbate the risk of catastrophic wildfire, with substantial impacts observed in the 21$^\text{st}$ century \citep{COVA2025123262, Iglesias2022fire}. In particular, severe wildfire can result in substantial losses of live tree AGB \citep{Jiang_2023}. Here, we define AGB loss to occur when live trees burn and carbon previously stored in aboveground biomass is redistributed among several post-fire pools and pathways, including immediate atmospheric emissions, transfer to standing and downed dead wood, incorporation into soil organic matter and pyrogenic carbon, and, in some cases, delayed losses during subsequent decomposition \citep{Loehman_2014, Santin_2015, DeLuca_2008}. In addition, severely burned areas may transition from positive live tree AGB before fire to zero live tree AGB after fire, which we define here as forest mortality. Although previous studies have examined wildfire impacts at smaller spatial scales \citep{CHEN2011140, boreal_forests_structure_changes_fire, CUI2025113889}, a comprehensive and comparable assessment of historical wildfire impacts on AGB across the West Coast has been lacking.

Statistical methods that incorporate auxiliary data and use sample observations from domains with greater data support to improve estimation in data-sparse domains, collectively referred to as small area estimation (SAE) techniques, have long been used in forest inventory modeling \citep{need_for_small_area_estimation}, including in wildfire-affected settings. In the interior western US, \cite{moisen_blackard_finco} combined sparse NFI data collected and maintained by the US Forest Service (USFS) Forest Inventory and Analysis (FIA) program with remotely sensed MODIS-based maps of forest presence and AGB to construct more precise estimates of forest AGB in small wildfire-affected areas using a composite SAE approach. By combining direct and indirect estimators, their approach balanced the instability of direct estimators based on only a few NFI plots with the potential bias of synthetic estimators derived primarily from remotely sensed maps. More recently, \cite{Gaines_Affleck_2021} compared two indirect SAE estimators for predicting forest regeneration following wildfire. In a spatio-temporal SAE setting, they demonstrated the value of borrowing information across both space and time to reduce domain-level uncertainty and highlighted the importance of accommodating spatial and temporal dependence when estimating wildfire impacts. \cite{affleck2023model} extended this work by developing a model-assisted estimator of postfire tree stocking. They achieved increases in precision over direct estimates, but observed instability in the bias correction term for sparsely sampled burned area domains. Collectively, these studies established the value of SAE methods for wildfire applications while also suggesting opportunities for more flexible model-based approaches that can accommodate complex spatio-temporal structures and provide richer uncertainty quantification for fine-scale inference.

Responding to calls for improved SAE methods in forest inventory applications \citep{Prisley2021, Cahoon2026}, recent methodological work has focused on adapting existing approaches and developing new ones to deliver fine-scale inference for forest attributes. These methods link high-resolution auxiliary information, typically from satellite or airborne remote sensing instruments, with sparse NFI plot data to produce more reliable estimates at fine spatial scales. Spatio-temporal SAE models further exploit dependence structures to share information across space and time. For example, \cite{Shannon_2025a,Shannon_2025b} developed spatio-temporal area-level SAE models for estimating county-level AGB across the contiguous US using remotely sensed percent tree canopy cover (TCC) data. Building on related developments in spatio-temporal SAE and hierarchical modeling \citep{pfeffermann2008, finley_et_al_2011}, \cite{may2025} introduced a zero-inflated spatio-temporal latent Gaussian model (LGM) for unit-level, point-referenced forest inventory data that accommodates the large number of zero-AGB observations arising from sampled nonforest conditions. To scale inference to large NFI datasets, the model uses a Nearest Neighbor Gaussian Process (NNGP; \citealt{datta_et_al_2016}) specification for the spatio-temporal random effects. This two-stage framework predicts forest presence/absence in the first stage and AGB conditional on forest presence in the second stage, while capturing short- and long-range spatio-temporal dependence commonly observed in forest attributes.

Building on this framework, we combine remotely sensed TCC data with plot-level measurements from the US NFI to quantify fine-scale AGB loss and forest mortality areas associated with wildfire. Our Bayesian spatio-temporal model provides spatially explicit inference on the magnitude and distribution of wildfire impacts on forest land across the West Coast of the US over the past two decades, together with corresponding uncertainty. We apply the model to 6,239 wildfires on National Forest System (NFS) lands in Washington, Oregon, and California from 2000 to 2022, providing a comprehensive regional assessment of wildfire-associated AGB loss, forest mortality area, and related derived quantities.

Our goal is to develop a consistent and flexible framework for quantifying both individual and aggregate wildfire impacts across the West Coast, while fully propagating uncertainty. Within the Bayesian framework, posterior distributions for derived quantities of interest arise naturally. In particular, we summarize and compare wildfire impacts using three primary quantities of interest:
\begin{itemize}
\item Total biomass loss (Mg), the total negative change in live tree AGB within a wildfire area, excluding instances of positive AGB change;
\item Forest mortality area (ha), area within a wildfire perimeter that transitions from positive live tree AGB before fire to zero live tree AGB after fire;
\item Wildfire severity (Mg/ha), average live tree AGB loss per unit forest mortality area as defined above.
\end{itemize}
We adopt the above structurally based definitions to align with the response variable, remote sensing measurements, modeling specification, and assumptions detailed in Section~\ref{sec:mod}.

The remainder of the paper is organized as follows. In Section~\ref{sec:methods}, we describe the data and model. In Section~\ref{sec:results}, we present model results and substantive findings. Finally, in Section~\ref{sec:discussion}, we offer discussion and concluding remarks.

\section{Methods}\label{sec:methods}

\subsection{Data}\label{sec:data}

NFI data are collected and published by the USFS FIA program for all states using a rotating panel design. Individual inventory plots are assigned to one of ten panels, with one panel visited each year, yielding a baseline 10-year remeasurement interval; however, some states and some ownerships are measured on shorter cycles \citep{betchold_patterson_2015}. 

We use FIA plot-level estimates of live tree AGB, derived from tree measurements, as the primary response data in the subsequent models \citep{fiadb_user_guide}. Each plot-year observation is represented by the exact FIA plot location and inventory year; repeated measurements of the same plot are retained as separate plot-year observations and accommodated in the subsequent models. To characterize live AGB status for the Bernoulli stage of the model, we classify inventory observations according to whether the plot-level live tree AGB estimate is positive or zero. This classification differs from FIA's forest/nonforest land status, which incorporates criteria such as minimum canopy cover, land use, evidence of recent tree cover, and capacity to return to forested conditions. As described in Section~\ref{sec:mod}, 90,547 plot-level FIA observations are used to fit the Bernoulli model for positive versus zero live AGB, while the subset of 43,951 positive live AGB observations is used to fit the LGM for square-root AGB.

\begin{figure}[ht!]
    \centering
    \includegraphics[width=8 cm]{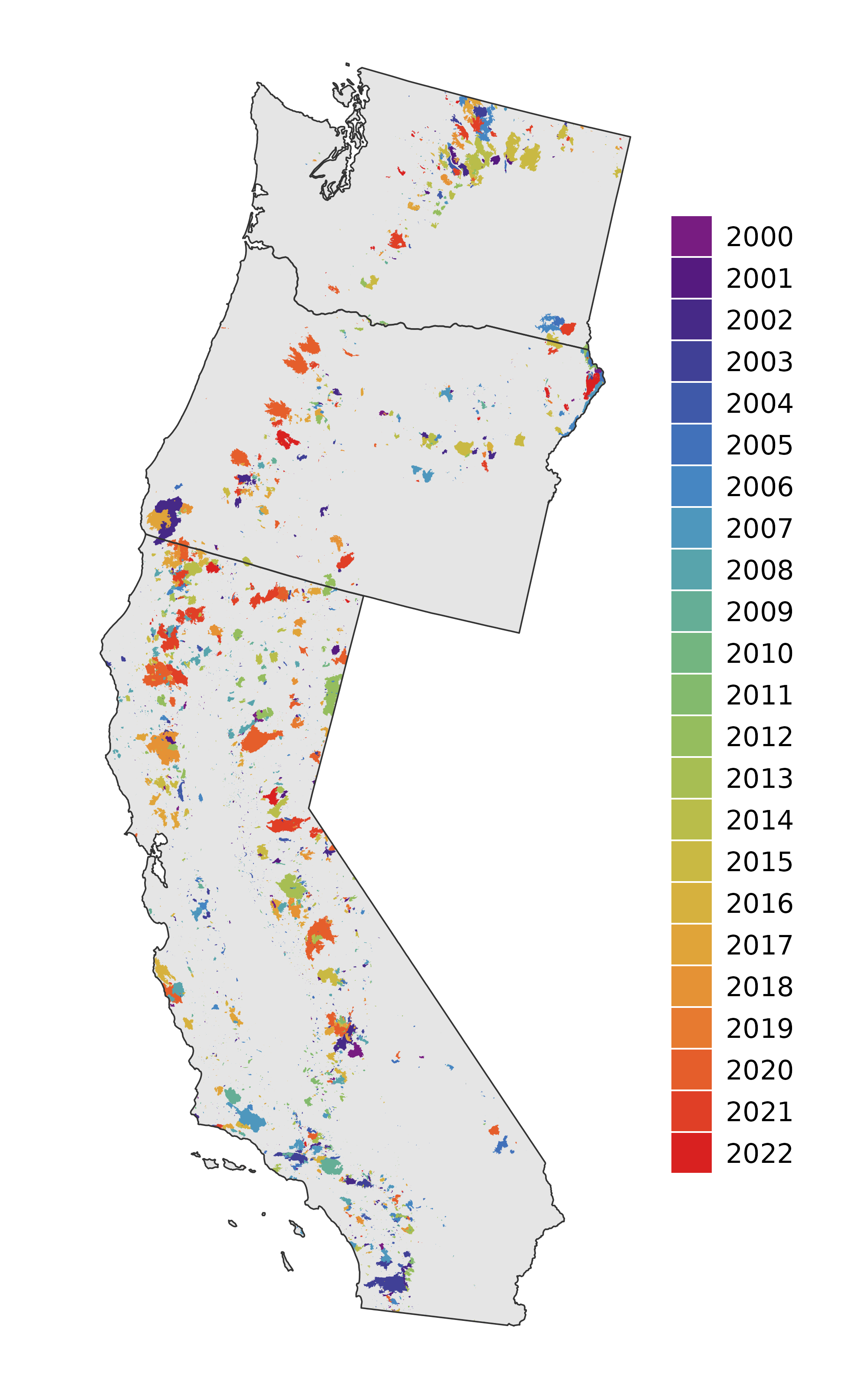}
    \caption{Wildland fire areas occurring on National Forest System lands in Washington, Oregon, and California from 2000 to 2022.}
    \label{fig:all_fires}
\end{figure}

Fire perimeters for wildland fires occurring on NFS lands are published by the USFS and include information such as fire year, stated cause, and a unique identifier. The dataset is intended to include fires larger than 4.05 ha (10 ac), although some smaller fires are also recorded. These data are used by forest managers to identify areas affected by wildfire, but they do not provide estimates of AGB loss or identify pixel-level forest mortality areas within mapped fire perimeters. We retain all recorded wildland fires larger than 0.405 ha (1 acre) occurring between 2000 and 2022 on NFS lands, yielding the 6,239 fires shown in Figure~\ref{fig:all_fires}. Same-year overlapping perimeters are filtered before fire-level and regional aggregation to avoid double counting, while fires in different years, including reburns, are retained as separate events. Although the perimeter dataset considered here includes only fires on NFS lands, it provides a comprehensive history of wildland fires on those landscapes and therefore supports consistent comparison over time. In addition, approximately 46\% of forested lands in Washington, Oregon, and California are managed by the NFS, providing substantial coverage of West Coast forest ecosystems \citep{usfs_data_discovery_2025}.

Satellite-based TCC raster data are produced by the USFS as part of the National Land Cover Database (NLCD), which estimates the percentage of land area covered by forest canopy using multispectral Landsat observations \citep{tcc_methods_2023}. These data are available as annual 30-by-30 m raster images from 1999 to 2023 and support fine-resolution comparison of forest structure over time. For model fitting, annual TCC is extracted from the raster cell containing each exact FIA plot location in the FIA inventory year. For prediction, TCC pixels within fire areas define the prediction locations, and TCC from the year preceding and the year following each fire is used to estimate pre- to post-fire changes in AGB and forest mortality area. We use TCC as the sole fixed-effect predictor because it is annual, spatially exhaustive, and closely aligned with the estimands.

\subsection{Model}\label{sec:mod}

We follow the zero-inflated spatio-temporal latent Gaussian modeling framework proposed by \cite{may2025}. The model proceeds in two stages: a Bernoulli data model for live forest presence or absence, followed by a LGM for square-root AGB conditional on live forest presence.

Let $y(\bbs, t) \geq 0$ denote the observed AGB density (Mg/ha) at an NFI plot at spatial location $\bbs$ and time $t$, where $\bbs$ denotes spatial coordinates and $t$ is measured in decimal years. To accommodate the substantial probability mass at zero in the AGB data, we define the binary process $p(\bbs, t)$ such that $p(\bbs, t)=1$ if $y(\bbs, t)>0$ and $p(\bbs, t)=0$ otherwise. The first stage of the model is then
\begin{align}
    p(\bbs, t) &\indsim \text{Bernoulli}(\mu_p(\bbs, t)), \\
    \text{logit}\left(\mu_p(\bbs, t)\right) &= \beta_{0,p} + x(\bbs, t)\beta_p,
    \label{eq:bernoulli_mod}
\end{align}
where $\beta_{0,p}$ is an intercept, $\beta_p$ is the regression coefficient associated with TCC, and $x(\bbs, t)$ denotes measured TCC at location $\bbs$ and time $t$.

Conditional on positive AGB, the second stage models the square-root transformed response $\tilde{y}(\bbs,t)=\sqrt{y(\bbs,t)}$ as
\begin{equation}
    \tilde{y}(\bbs, t) = \beta_{0,y} + x(\bbs, t)\beta_y + u_y(\bbs, t) + \varepsilon_y(\bbs, t),
    \label{eq:carbon_mod}
\end{equation}
where $\varepsilon_y(\bbs, t) \iidsim \text{Normal}(0, \tau^2)$ is an independent residual error term, $\beta_{0,y}$ is an intercept, $\beta_y$ is the regression coefficient associated with $x(\bbs, t)$, and $u_y(\bbs, t)$ is a mean-zero spatio-temporal random effect. The Bernoulli model is fit using all FIA observations, with positive live AGB observations coded as 1 and zero live AGB observations coded as 0, whereas the square-root AGB model is fit only to the subset of observations with positive observed AGB. While the Bernoulli model could include a spatial random effect (e.g., $u_p(\bbs, t)$ in \eqref{eq:bernoulli_mod}), our exploratory analysis showed that TCC captured the majority of the spatio-temporal structure in live forest occurrence, removing the need for an additional residual process term.

The spatio-temporal process $u_y(\bbs,t)$ is specified as a sum of independent Gaussian process (GP) components, following \cite{may2025}. These components allow variation to be represented at different spatial and temporal scales. Here, we use two components to capture short- and long-range spatio-temporal dependence, which has been shown to represent local- and landscape-scale patterns in AGB effectively. Specifically, let $u_y(\bbs, t) \sim GP(0, C(\cdot,\cdot))$, where $C(\delta_s,\delta_t)$ is a covariance function depending on spatial lag $\delta_s$ and temporal lag $\delta_t$. The two-component covariance structure is
\begin{equation}
    C(\delta_s, \delta_t) = \sigma_y^2 \left( (1-\alpha_y) e^{-\delta_s\phi_{y,1} - \delta_t\lambda_{y,1}} + \alpha_y e^{-\delta_s\phi_{y,2} - \delta_t\lambda_{y,2}} \right),
    \label{eq:GP_components}
\end{equation}
where $\phi_{y,1}>0$ and $\lambda_{y,1}>0$ are the spatial and temporal decay parameters for the local-scale component, and $\phi_{y,2}>0$ and $\lambda_{y,2}>0$ are the corresponding decay parameters for the landscape-scale component. Identifiability between the two components is enforced through the prior distributions on $\phi_{y,1}$ and $\phi_{y,2}$. Further, $\sigma_y^2>0$ is the marginal spatio-temporal variance parameter, and the relative contribution of the two components is controlled by $0<\alpha_y<1$, with $\alpha_y$ weighting the landscape-scale component. The full model therefore includes regression parameters $\beta_{0,p}$, $\beta_p$, $\beta_{0,y}$, and $\beta_y$; spatio-temporal covariance parameters $\sigma_y^2$, $\alpha_y$, $\phi_{y,1}$, $\phi_{y,2}$, $\lambda_{y,1}$, and $\lambda_{y,2}$; and residual variance parameter $\tau^2$.

Given the substantial size of the spatio-temporal dataset considered here, direct GP inference of $u_y(\bbs, t)$ is computationally infeasible. In particular, Cholesky factorization of the full spatio-temporal covariance matrix has cubic computational complexity in the number of observations. We therefore replace the GP with its corresponding 15-neighbor NNGP approximation, a neighbor size shown to approximate full GP inference well in large spatial datasets \citep{datta_et_al_2016, Finley_nngp}.

The Bayesian model specification is completed by assigning prior distributions to all parameters. Regression coefficients in the Bernoulli stage are assigned independent Normal priors with mean 0 and standard deviation 10, while regression coefficients in the Gaussian stage are assigned flat priors. Variance parameters are assigned inverse-Gamma priors, and uniform priors are assigned to the remaining covariance parameters, with $\alpha_y \sim U(0,1)$, $\phi_{y,1} \sim U(0.3,30)$, $\phi_{y,2} \sim U(0.001,0.29)$, $\lambda_{y,1} \sim U(0.001,3)$, and $\lambda_{y,2} \sim U(0.001,3)$. Distance is measured in kilometers. Under the exponential correlation function, the effective range is approximately the distance or time lag at which the correlation drops to 0.05. Thus, the priors on the spatial decay parameters imply effective spatial range support of approximately 0.1 to 10 km for the local-scale component and 10 to 3000 km for the landscape-scale component, while the priors on the temporal decay parameters imply effective temporal range support of approximately 1 to 3000 years for both components. Supplementary Material S1 provides a prior-support sensitivity analysis showing that moving the spatial split separating the two covariance components from 10 to 50 km effective range had negligible impact on fire-level posterior summaries and rankings.

Inference for model parameters and predictions is based on Markov chain Monte Carlo (MCMC) samples from the posterior and posterior predictive distributions. A P{\'o}lya--Gamma data-augmentation sampler is used to update the Bernoulli model regression coefficients through Gibbs sampling \citep{Polson2013}. Full conditional distributions are also sampled via Gibbs updates for the LGM regression coefficients, while Metropolis updates are used for the variance parameters, $\alpha_y$, and the decay parameters. We used four MCMC chains of 30,000 iterations, discarding the first 15,000 iterations from each chain for parameter inference and convergence diagnostics. Fire-pixel prediction retained 100 evenly spaced posterior predictive draws from one post-burn-in chain. 

For a generic prediction pixel $\bbs_0$ and time $t_0$, the Gaussian stage provides a posterior predictive draw $\tilde{y}(\bbs_0, t_0)$ on the square-root AGB scale. We transform each draw back to Mg/ha as $y(\bbs_0, t_0)=\tilde{y}(\bbs_0, t_0)^2$ and form the corresponding zero-inflated AGB draw as $z(\bbs_0, t_0) = p(\bbs_0, t_0)y(\bbs_0, t_0)$. Because the back transformation is applied directly to each posterior predictive draw, no separate retransformation-bias adjustment is needed \citep[e.g.,][]{Stow2006}. The zero-inflated AGB predictive distribution for $z(\bbs_0, t_0)$, and subsequent derived quantities of interest defined in Section~\ref{sec:derived_quantities}, are computed sample-for-sample from the paired posterior predictive draws of $p(\bbs_0, t_0)$ and $y(\bbs_0, t_0)$. 

Fire-, year-, and state-level summaries are computed within each draw, and reported credible intervals are posterior quantiles of those draw-level summaries. These intervals are conditional on the fitted model, prior distributions, FIA observations and exact plot locations, TCC rasters, fire perimeters, and annual temporal alignment.

All model fitting, prediction, and posterior summarization were implemented in R using the \texttt{stLMM} package \citep{stLMM}. The analysis scripts are included as supplementary material, and Supplementary Material S1 provides an overview of how those scripts fit the Bernoulli and Gaussian model stages, generate fire-pixel posterior predictions, combine the zero-inflated posterior predictive samples, run diagnostics and the prior-support sensitivity analysis, and reproduce the reported tables and figures. Because exact FIA plot locations cannot be redistributed, the scripts are provided without FIA plot data. The full manuscript results are reproducible with the provided scripts given access to the FIA, TCC, and fire-perimeter input data used in the analysis.

\subsection{Derived Quantities}\label{sec:derived_quantities}

The model described in Section~\ref{sec:mod} is fit to the plot-level FIA AGB data described in Section~\ref{sec:data}. Predictions are then obtained at the level of TCC pixels, which constitute the fundamental prediction units in our analysis. For a fire occurring in year $t_0$, let $\mathcal{S}_0$ denote the set of prediction-pixel locations within the mapped fire area. For each prediction pixel $\bbs_0 \in \mathcal{S}_0$, posterior predictive samples are available for $p(\bbs_0, t_0-1)$, $p(\bbs_0, t_0+1)$, $z(\bbs_0, t_0-1)$, and $z(\bbs_0, t_0+1)$. We use these pre- and post-fire predictions to compute total AGB loss, forest mortality area, and wildfire severity. As noted in Section~\ref{sec:intro}, we adopt specific definitions of these metrics that are consistent with the modeling specification and assumptions outlined in Section~\ref{sec:mod}.

Let $a$ denote the area of a TCC pixel in hectares. For the 30-by-30 m raster used here, $a = 0.09$ ha.

For a prediction pixel $\bbs_0$ experiencing wildfire in year $t_0$, forest mortality is identified when $d(\bbs_0, t_0) = p(\bbs_0, t_0 - 1) - p(\bbs_0, t_0+1) = 1$, or in other words, when $\bbs_0$ is classified as transitioning from positive live AGB before fire to zero live AGB after fire. For a given posterior predictive draw, let $\mathcal{D}_0(t_0) = \{\bbs_0 \in \mathcal{S}_0: d(\bbs_0,t_0)=1\}$ denote the subset of prediction pixels experiencing forest mortality. The total forest mortality area is calculated as
\begin{equation}\label{eq:deforestation}
d(\mathcal{S}_0, t_0) = a \sum_{\bbs_0 \in \mathcal{S}_0} \indicator_{d(\bbs_0,t_0) = 1},
\end{equation}
where $\indicator$ is the indicator function.

Similarly, the AGB loss for a prediction pixel $\bbs_0$ experiencing forest mortality associated with wildfire in year $t_0$ is $a\left(z(\bbs_0,t_0-1) - z(\bbs_0,t_0+1)\right) = a z(\bbs_0,t_0-1)$, since $z(\bbs_0, t_0+1) = 0$ for that pixel and draw. For the set of locations $\bbs_0 \in \mathcal{D}_0(t_0) \subset \mathcal{S}_0$ experiencing forest mortality, the total AGB loss in these areas is calculated as
\begin{equation}\label{eq:bio_loss}
l(\mathcal{D}_0(t_0), t_0) = a \sum_{\bbs_0 \in \mathcal{D}_0(t_0)} z(\bbs_0, t_0-1).
\end{equation}

Wildfire severity for wildfire area $\mathcal{S}_0$ in year $t_0$ is summarized as the average AGB loss per unit forest mortality area, and is directly available as $l(\mathcal{D}_0(t_0), t_0) / d(\mathcal{S}_0, t_0)$. We use forest mortality area rather than total mapped fire area in the denominator so that wildfire severity summarizes biomass-loss intensity among pixels predicted to transition from positive live AGB before fire to zero live AGB after fire. If $d(\mathcal{S}_0, t_0)=0$ for a posterior predictive draw, then $\mathcal{D}_0(t_0)$ is empty and $l(\mathcal{D}_0(t_0), t_0)=0$, so wildfire severity is undefined for that draw. These undefined draw-level severity values are omitted from posterior summaries of severity; the corresponding fire remains included in total AGB loss and forest mortality area summaries.

Finally, total AGB loss for all locations $\bbs_0 \in \mathcal{S}_0$ within a given wildfire area in year $t_0$ is calculated as
\begin{equation}\label{eq:total_loss}
l(\mathcal{S}_0, t_0) = a \sum_{\bbs_0 \in \mathcal{S}_0} \left[z(\bbs_0, t_0+1) - z(\bbs_0, t_0-1)\right] \indicator_{z(\bbs_0, t_0+1) - z(\bbs_0, t_0-1) < 0}.
\end{equation}

In this way, we summarize negative AGB changes within wildfire areas while excluding positive AGB changes within fire perimeters that would otherwise offset estimated losses. Further, by considering wildfire impacts at the pixel level, we rely less on the accuracy of reported wildfire boundaries and can facilitate direct comparison among fires in different ecosystems over time. Although $l(\mathcal{S}_0, t_0)$ is defined as a signed negative change, we report its magnitude when summarizing wildfire-associated AGB loss.

\section{Results}\label{sec:results}

The model described in Section~\ref{sec:mod} was fit to the plot-level NFI AGB data described in Section~\ref{sec:data}. Posterior summaries for model parameters are given in Table~\ref{tab:posterior_params}. Derived quantities of interest described in Section~\ref{sec:derived_quantities}, including forest mortality area, AGB loss, and wildfire severity, were obtained for pixel locations within mapped wildfire areas from posterior predictive samples and then aggregated to the levels of individual fires and states for comparison.

\begin{table}[ht!]
\caption{Posterior summaries for model parameters described in Section~\ref{sec:mod}. Estimates are given as posterior medians with 95\% credible intervals in parentheses.}
\begin{center}
\begin{tabular}{lc}
\toprule
Parameter & Posterior Estimate\\
\midrule
$\beta_{0,p}$ & -2.217 (-2.246, -2.189)\\
$\beta_p$ & 0.083 (0.082, 0.084)\\
$\beta_{0,y}$ & 1.224 (1.145, 1.302)\\
$\beta_y$ & 0.061 (0.060, 0.062)\\
$\sigma_y^2$ & 3.144 (3.071, 3.219)\\
$\alpha_y$ & 0.230 (0.213, 0.247)\\
$\phi_{y,1}$ & 8.368 (4.030, 26.866)\\
$\phi_{y,2}$ & 0.038 (0.031, 0.045)\\
$\lambda_{y,1}$ & 0.0112 (0.0098, 0.0124)\\
$\lambda_{y,2}$ & 0.0119 (0.0100, 0.0140)\\
$\tau^2$ & 0.075 (0.053, 0.107)\\
\bottomrule
\end{tabular}
\label{tab:posterior_params}
\end{center}
\end{table}

Posterior summaries in Table~\ref{tab:posterior_params} indicate expected positive relationships between percent TCC and both live forest presence and AGB, as reflected by $\beta_p$ and $\beta_y$, respectively. Specifically, larger TCC values are associated with higher posterior probability of positive live AGB and, conditional on positive live AGB, larger square-root AGB. Following the two-component covariance structure in \eqref{eq:GP_components}, posterior estimates for the local- and landscape-scale spatial decay parameters imply effective spatial ranges of approximately 0.4 km and 78.64 km, respectively. Thus, the two components capture residual AGB patterns at fine within-stand scales and broader landscape scales after accounting for TCC. Posterior estimates for $\lambda_{y,1}$ and $\lambda_{y,2}$ imply effective temporal ranges of approximately 266.5 and 252.5 years, respectively, reflecting gradual temporal change in residual forest structure (i.e., after accounting for variance explained by TCC). The posterior median of $\alpha_y$ is 0.230, which corresponds to approximately 23.0\% of the marginal spatio-temporal variance assigned to the landscape-scale component and 77.0\% assigned to the local-scale component.

We estimated forest mortality associated with all 6,239 fires occurring between 2000 and 2022 on NFS lands in Washington, Oregon, and California. In total, an estimated 95,260,702 Mg (94,700,728, 95,865,692) of AGB were lost within wildfire areas, where parentheses here and throughout denote 95\% credible intervals. In addition, 2,391,984 ha (2,388,519, 2,394,977) were estimated to have transitioned from positive live AGB before fire to zero live AGB after fire during the study period. These forest mortality areas were associated with 54,145,581 Mg (53,637,460, 54,555,488) of AGB loss, resulting in an average wildfire severity of 22.63 Mg/ha (22.44, 22.77) of forest mortality area.

\begin{figure}[ht!]
    \centering
    %[trim={left bottom right top},clip]
    \includegraphics[trim={4cm 0cm 4cm 1.3cm},clip,width=10 cm]{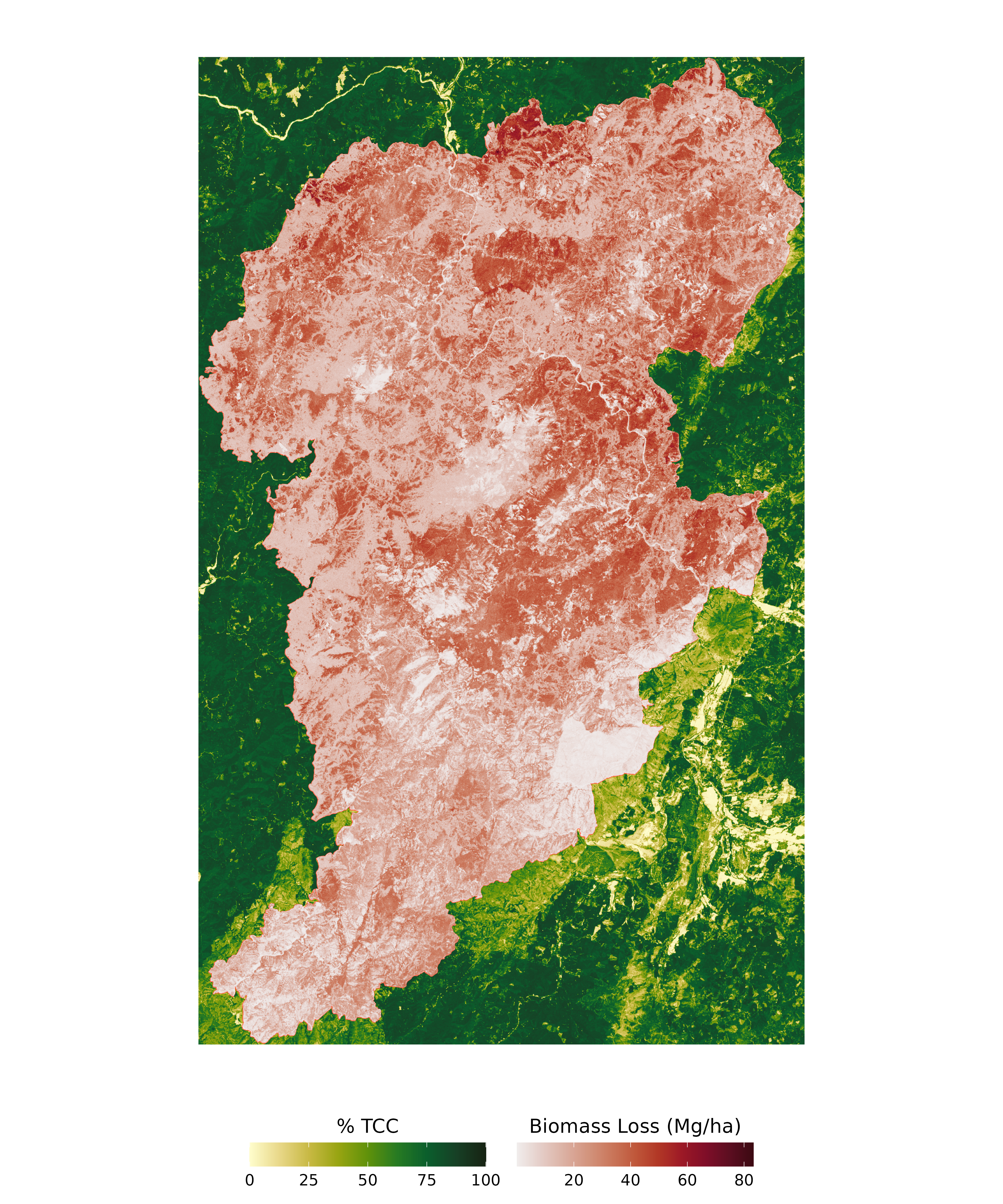}
    \caption{Pixel-level posterior mean total AGB loss for the Biscuit Fire, which burned in southwestern Oregon in 2002.}
    \label{fig:biscuit_biomass}
\end{figure}

\begin{figure}[ht!]
    \centering
    %[trim={left bottom right top},clip]
    \includegraphics[trim={3cm 0cm 3cm 1.7cm},clip,width=10 cm]{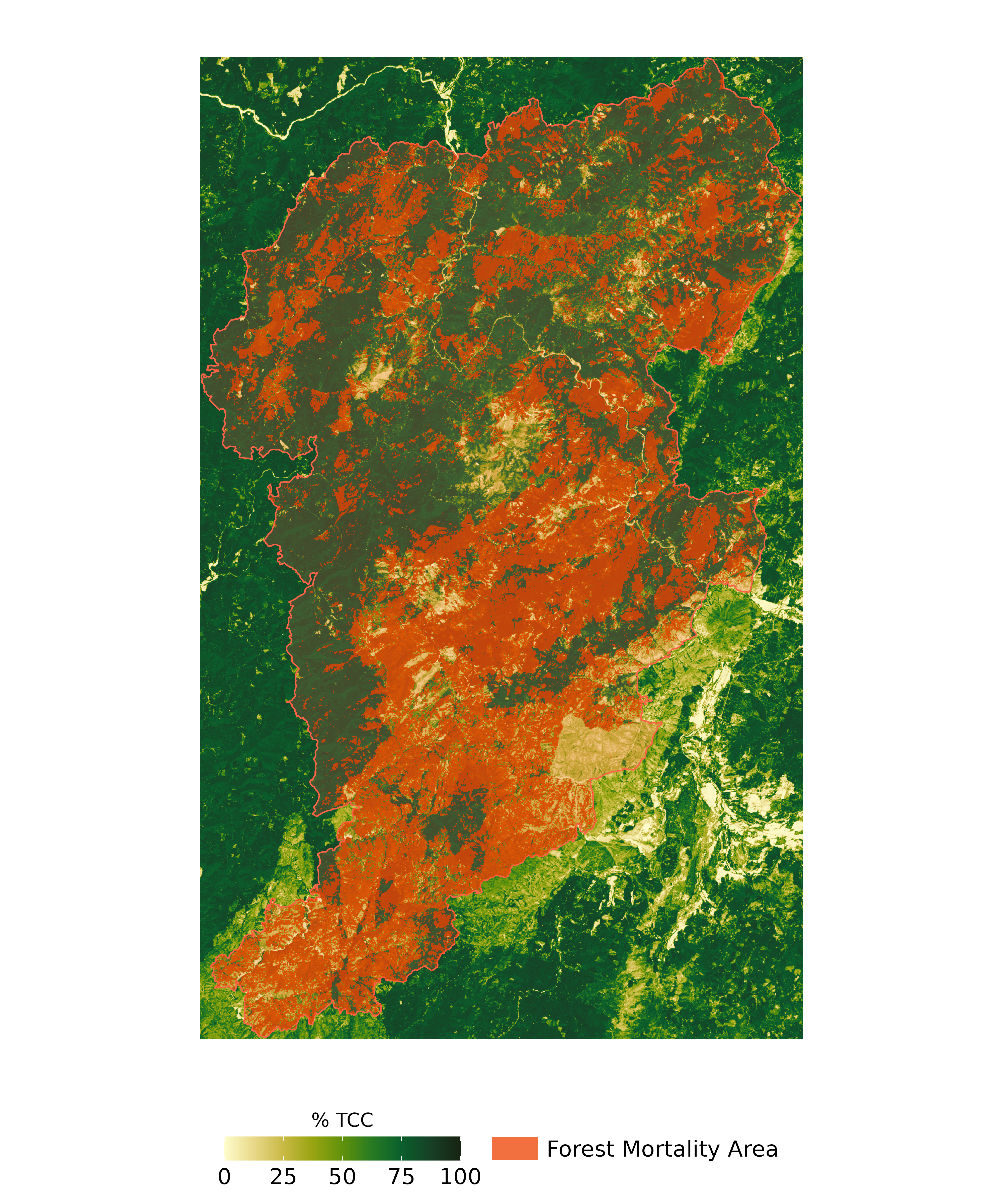}
    \caption{Pixel-level forest mortality area associated with the Biscuit Fire. For mapping purposes, pixels were classified as forest mortality areas if their posterior predictive probability of having positive AGB in year $t-1$ and zero AGB in year $t+1$, for a fire occurring in year $t$, exceeded 0.5. The Biscuit Fire resulted in the greatest total forest mortality area and the greatest AGB loss in areas experiencing forest mortality among all fires examined.}
    \label{fig:biscuit_deforest}
\end{figure}

\begin{figure}[ht!]
    \centering
    %[trim={left bottom right top},clip]
    \includegraphics[trim={2cm 0cm 2cm 1cm},clip, width=10 cm]{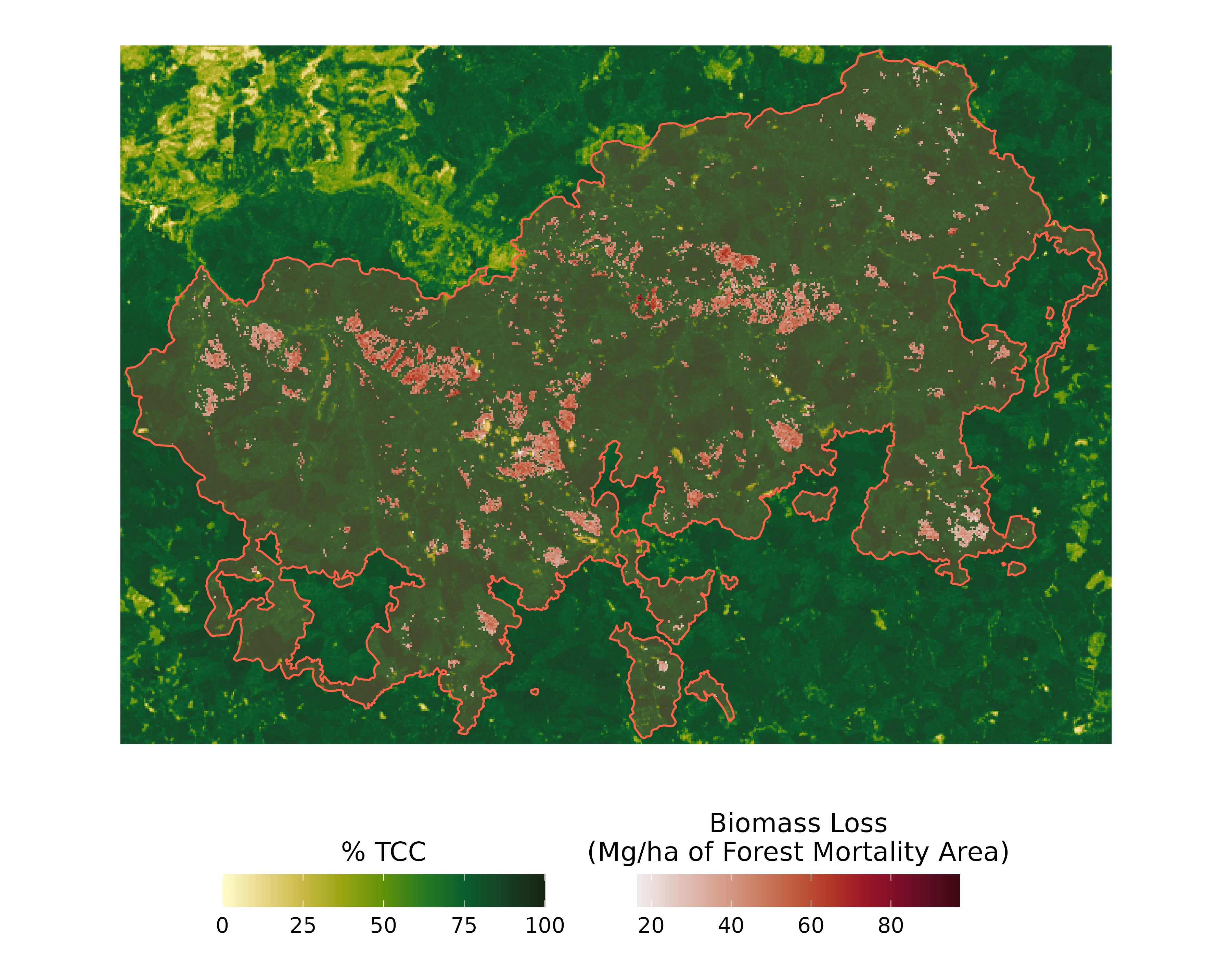}
    \caption{Pixel-level posterior mean AGB loss for forest mortality areas in the Gales Fire, which burned in the Willamette National Forest in Oregon. For mapping purposes, pixels were classified as forest mortality areas if their posterior predictive probability of having positive AGB in year $t-1$ and zero AGB in year $t+1$, for a fire occurring in year $t$, exceeded 0.5. The Gales Fire is shown as an example of high wildfire severity with substantial estimated forest mortality area.}
    \label{fig:gales_fire}
\end{figure}

Looking across individual fires, the 2002 Biscuit Fire in southwestern Oregon resulted in the greatest total AGB loss, estimated at 4,469,920 Mg (4,423,774, 4,537,477) (Figure~\ref{fig:biscuit_biomass}). The Biscuit Fire also resulted in the greatest mortality area, totaling 79,230 ha (78,881, 79,591) (Figure~\ref{fig:biscuit_deforest}). The Gales Fire provides an example of high wildfire severity with substantial estimated forest mortality area (Figure~\ref{fig:gales_fire}), and was associated with an average AGB loss of 70.16 Mg/ha (68.71, 72.03) of forest mortality area in the Willamette National Forest in Oregon.

\begin{figure}[ht!]
    \centering
    %[trim={left bottom right top},clip]
    \includegraphics[trim={1.2cm 1.5cm 0cm 2cm},clip, width=\textwidth]{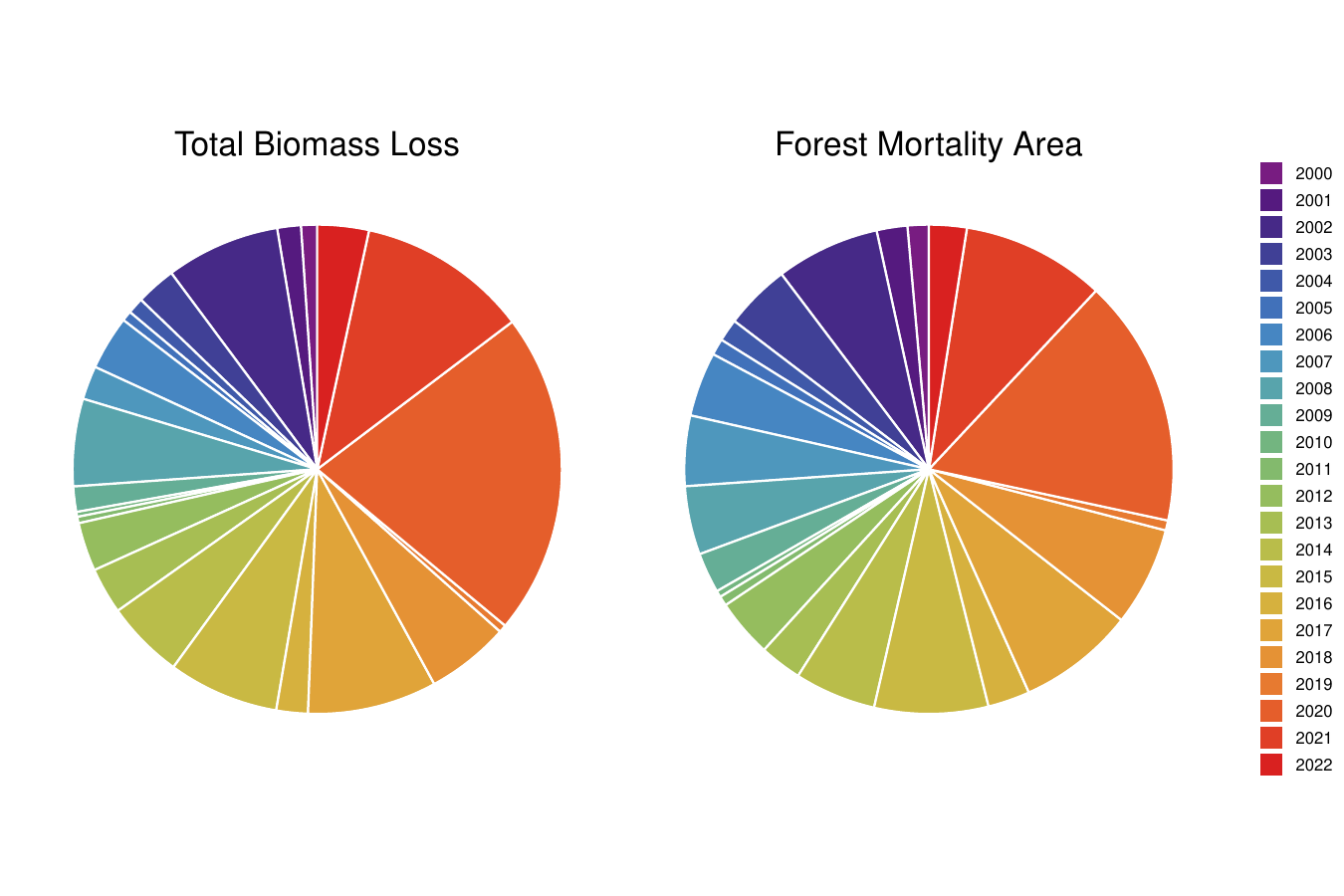}
    \caption{Annual comparison of total AGB loss and forest mortality area.}
    \label{fig:pie_charts}
\end{figure}

\begin{figure}[ht!]
    \centering
    %[trim={left bottom right top},clip]
    \includegraphics[width=\textwidth]{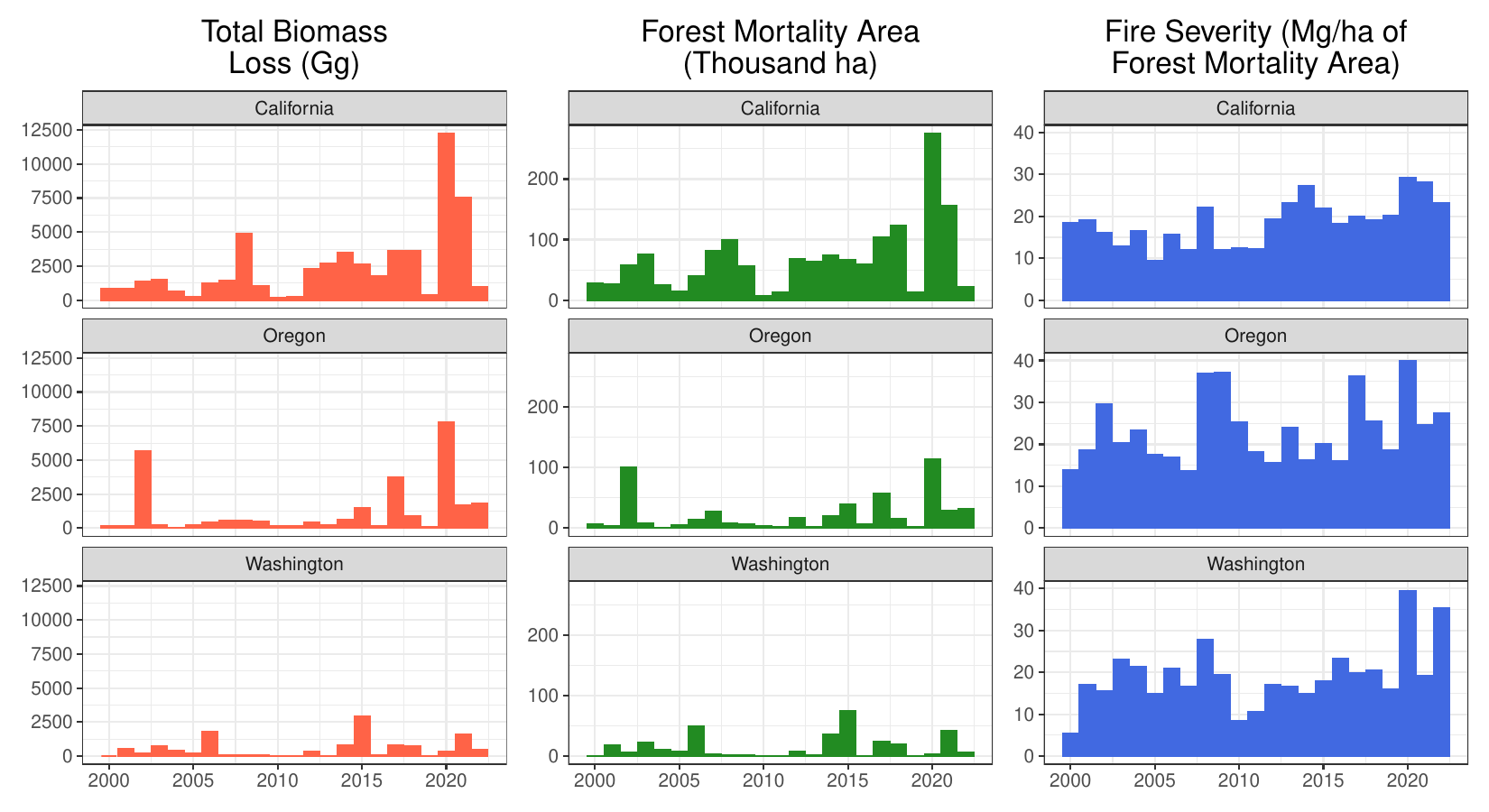}
    \caption{State-level summaries of posterior median total AGB loss, forest mortality area, and wildfire severity between 2000 and 2022 for Washington, Oregon, and California.}
    \label{fig:state_summaries}
\end{figure}

During the study period, both total AGB loss and forest mortality area were greatest in 2020 (Figure~\ref{fig:pie_charts}), and recent years in the last decade generally exhibited greater forest mortality area and AGB loss than earlier years. At the state level, California exhibited the greatest total AGB loss and forest mortality area, and wildfire severity was greatest in Oregon (Figure~\ref{fig:state_summaries}).

\section{Discussion}\label{sec:discussion}

We implemented a zero-inflated Bayesian spatio-temporal SAE model using NFI and remotely sensed TCC data to quantify AGB loss and forest mortality area associated with more than 20 years of wildfire on the West Coast of the US. Pixel-level AGB estimates before and after each wildfire were obtained, along with derived quantities of interest including wildfire severity. Estimates were also aggregated from the pixel level to individual wildfires and states to identify broader spatial patterns and trends in wildfire impacts.

In total, an estimated 95,260,702 Mg of AGB were lost due to wildfire between 2000 and 2022, with an associated 2,391,984 ha of forest mortality area. Relative to pre-fire forest AGB and live forest area within published wildfire boundaries, these estimates represent a loss of 67.17\% (66.95, 67.37) of total AGB and 46.58\% (46.35, 46.82) of pre-fire live forest area transitioning to zero AGB. By evaluating pixel-level changes in live forest conditions using NFI plot measurements and remotely sensed TCC data, our estimates of forest mortality area and AGB loss reflect within-fire changes in forest structure rather than simple summaries of total fire-perimeter area. Importantly, the results highlight the uneven distribution of forest mortality area and AGB loss within reported wildfire boundaries, as illustrated in Figures~\ref{fig:biscuit_deforest} and \ref{fig:gales_fire}. Further, AGB losses in forest mortality areas amounted to 54,145,581 Mg, resulting in an average wildfire severity of 22.63 Mg/ha of forest mortality area.

Comparing total AGB loss and forest mortality area over the study period (Figure~\ref{fig:pie_charts}), 2020 stands out as the most consequential year for both outcomes. A number of factors contributed to the historic 2020 fire season on the West Coast. In the Pacific Northwest, several severe wildfires occurring around Labor Day were driven by strong east winds and unusually dry conditions \citep{hcascadia_2020_wildfires}. Importantly, the 2020 Pacific Northwest fires included unusually extensive burning in forests west of the Cascades, which burn relatively infrequently. Similarly, California experienced a record-breaking number of large wildfires in 2020 \citep{2020_california_fires}. These fires were influenced in part by substantial drought conditions during the decade preceding the 2020 fire season. Together, these factors, along with the growing influence of climate change, contributed to some of the most consequential fires observed on the West Coast in the last century \cite{higuera2021record}.

Examining individual fires during the study period, posterior estimates of total AGB loss, forest mortality area, and wildfire severity provide fire-level summaries that further highlight the distinct characteristics of forest regions across the West Coast. The fire associated with the greatest estimated total AGB loss was the 2002 Biscuit Fire in southwestern Oregon, which resulted in 4,469,920 Mg of total AGB loss and 79,230 ha of forest mortality area. Importantly, the pixel-level estimation approach used here highlights the uneven impact of the Biscuit Fire within the published wildfire boundary, which is consistent with previous work examining the effects of the fire across stands of different ages and species classes \citep{biscuit_fire}.

Finally, we consider wildfire severity as the average amount of AGB lost in forest mortality areas. The Gales Fire in Oregon's Willamette National Forest provides an example of a high-severity fire with substantial estimated forest mortality area. Specifically, the Gales Fire was associated with 70.16 Mg/ha of AGB loss in forest mortality areas (Figure~\ref{fig:gales_fire}). Although the Gales Fire was not especially large, the burned portions of its perimeter were associated with high AGB density. Fires such as the Gales Fire likely contribute to the pattern shown in Figure~\ref{fig:state_summaries}, where the high-AGB forests of Washington and Oregon exhibit wildfire severities comparable to, or greater than, those of California.

The two-stage spatio-temporal LGM approach used here to characterize pixel-level wildfire impacts across the recent history of West Coast fires provides opportunities to better understand wildfire-impact patterns and to aggregate results across space and time. However, several limitations and opportunities for future development remain. First, opportunities to validate estimated AGB loss and forest mortality area are limited, because fine-resolution reference data on wildfire impacts are not readily available. However, the LGM used here has previously been applied in AGB estimation studies using NFI and remote sensing data \citep{Shannon2025_biorxiv, may2025}. In particular, \cite{may2025} conducted a cross-validation study for the proposed model and reported superior performance relative to a submodel with a single-component covariance structure. Second, only wildfires recorded on NFS lands are considered here, which may influence aggregated annual and state summaries. However, as noted earlier, the wildfire perimeter dataset provides a consistent view of West Coast forest ecosystems over the study period. Future work could extend this analysis using additional wildfire datasets. A third limitation is that changes in AGB within wildfire perimeters may reflect compound disturbance effects rather than fire alone. Drought, bark beetle outbreaks, other insect and disease agents, and forest management activities can all contribute to tree mortality and changes in live AGB within the same spatial and temporal window considered here. Thus, although we refer to these changes as wildfire-associated because they occur within mapped wildfire perimeters between the pre- and post-fire prediction years, they should not always be interpreted as purely fire-caused. Future work could address this limitation by incorporating additional disturbance and management information, such as USFS Aerial Detection Survey products for insect, disease, and drought-related mortality, Landscape Change Monitoring System products that map vegetation loss, vegetation gain, land cover, and land use change processes, and emerging time series from missions such as NISAR, whose high-resolution L-band SAR observations and 12-day repeat cycle could help distinguish rapid fire effects from other disturbance and recovery processes \citep{nisar_handbook_2019}.

Historical wildfire-associated changes in West Coast forest AGB loss and forest mortality area have been substantial, and models that provide comprehensive pixel-level estimation of these changes with associated uncertainty are critical for understanding trends and patterns across broad spatial and temporal domains. Previous studies have identified parts of the West Coast as potential carbon sources due to wildfire \citep{Shannon_2025a, Shannon_2025b}, and the fine-resolution analysis developed here provides a more complete picture of wildfire-associated forest change over time. Under a Bayesian framework, posterior estimates for primary and derived quantities of interest, including AGB and forest mortality area, are equipped with full uncertainty quantification. This stands in contrast to broad summaries of area burned based solely on wildfire perimeters (see, e.g., \citep{Iglesias2022fire}), which may not reflect within-fire patterns of burn severity and forest change. Because these summaries are expressed in biomass units, they can be propagated more directly into carbon-accounting calculations than spectral burn-severity indices. The posterior predictive distribution also helps distinguish large, well-supported biomass-loss patterns from apparent differences that are small relative to uncertainty. Thus, the framework is complementary to conventional burn-severity products: those products describe spectral vegetation change, whereas the estimates here quantify inventory-informed structural change. More broadly, we provide a flexible modeling framework that can scale to a variety of spatial domains and data sources for future attribution studies of forest AGB dynamics.

\clearpage
\bibliographystyle{apalike}
\bibliography{literature}

\begin{thebibliography}{}

\bibitem[Affleck and Gaines, 2023]{affleck2023model}
Affleck, D.~L. and Gaines, III, G.~C. (2023).
\newblock Model-assisted domain estimation of postfire tree regeneration in the
  western {US} using nearest neighbor techniques.
\newblock {\em Canadian Journal of Forest Research}, 53(12):981--995.

\bibitem[Aznar-Sánchez et~al., 2018]{sanchez_ecosystem_services}
Aznar-Sánchez, J.~A., Belmonte-Ureña, L.~J., López-Serrano, M.~J., and
  Velasco-Muñoz, J.~F. (2018).
\newblock {Forest Ecosystem Services}: An analysis of worldwide research.
\newblock {\em Forests}, 9(8).

\bibitem[Azuma et~al., 2004]{biscuit_fire}
Azuma, D.~L., Donnegan, J., and Gedney, D. (2004).
\newblock {\em {Southwest Oregon} {Biscuit Fire}: an analysis of forest
  resources and fire severity}.
\newblock Res. Pap. PNW-RP-560. Portland, OR: U.S. Department of Agriculture,
  Forest Service, Pacific Northwest Research Station.

\bibitem[Betchold and Patterson, 2015]{betchold_patterson_2015}
Betchold, W.~A. and Patterson, P.~L. (2015).
\newblock {\em The Enhanced {Forest Inventory and Analysis} Program -- National
  Sampling Design and Estimation Procedures}.
\newblock Gen. Tech. Rep. SRS-80. Asheville, NC: U.S. Department of Agriculture
  Forest Service, Southern Research Station.

\bibitem[Cahoon et~al., 2026]{Cahoon2026}
Cahoon, S.~M., Randolph, K., Gaines, III, G.~C., Crocker, S., Thomas, V., and
  Munro, H. (2026).
\newblock Advancing forest inventory science for enhanced insights into
  tomorrow's forests.
\newblock {\em Forest Ecology and Management}, 602:123405.

\bibitem[Chen et~al., 2011]{CHEN2011140}
Chen, X., Liu, S., Zhu, Z., Vogelmann, J., Li, Z., and Ohlen, D. (2011).
\newblock Estimating aboveground forest biomass carbon and fire consumption in
  the {U.S.} {U}tah {H}igh {P}lateaus using data from the {Forest Inventory and
  Analysis} program, {Landsat}, and {LANDFIRE}.
\newblock {\em Ecological Indicators}, 11(1):140--148.

\bibitem[Cova et~al., 2025]{COVA2025123262}
Cova, G.~R., Prichard, S.~J., Zald, H.~S., Gaines, W.~L., and Kane, V.~R.
  (2025).
\newblock Implications of recent wildfires for forest management on federal
  lands in the {Pacific Northwest}, {USA}.
\newblock {\em Forest Ecology and Management}, 598:123262.

\bibitem[Cui et~al., 2025]{CUI2025113889}
Cui, L., Xu, X., and Chen, S. (2025).
\newblock Estimating biomass consumption and carbon emissions by integrating
  {GEDI}, {Sentinel-1}, and {Sentinel-2} data.
\newblock {\em Ecological Indicators}, 178:113889.

\bibitem[Datta et~al., 2016]{datta_et_al_2016}
Datta, A., Banerjee, S., Finley, A.~O., and Gelfand, A.~E. (2016).
\newblock Hierarchical {Nearest-Neighbor} {Gaussian} process models for large
  geostatistical datasets.
\newblock {\em Journal of the American Statistical Association},
  111(514):800--812.

\bibitem[DeLuca and Aplet, 2008]{DeLuca_2008}
DeLuca, T.~H. and Aplet, G.~H. (2008).
\newblock Charcoal and carbon storage in forest soils of the {Rocky Mountain
  West}.
\newblock {\em Frontiers in Ecology and the Environment}, 6(1):18--24.

\bibitem[Feng et~al., 2024]{boreal_forests_structure_changes_fire}
Feng, T., Duncanson, L., Hancock, S., Montesano, P., Skakun, S., Wulder, M.~A.,
  White, J.~C., Minor, D., and Loboda, T. (2024).
\newblock Characterizing fire-induced forest structure and aboveground biomass
  changes in boreal forests using multitemporal {LiDAR} and {Landsat}.
\newblock {\em IEEE Journal of Selected Topics in Applied Earth Observations
  and Remote Sensing}, 17:10108--10125.

\bibitem[{FIADB}, 2025]{fiadb_user_guide}
{FIADB} (2025).
\newblock {\em The {Forest Inventory and Analysis} Database User Guide
  ({NFI})}.
\newblock {U.S.} Department of Agriculture, Forest Service.
\newblock Last updated August 25, 2025.

\bibitem[Finley et~al., 2026]{stLMM}
Finley, A.~O., Banerjee, S., Datta, A., and May, P.~B. (2026).
\newblock {\em {stLMM}: Bayesian Spatial and Space-Time Linear Mixed Models}.
\newblock R package.

\bibitem[Finley et~al., 2011]{finley_et_al_2011}
Finley, A.~O., Banerjee, S., and MacFarlane, D.~W. (2011).
\newblock A hierarchical model for quantifying forest variables over large
  heterogeneous landscapes with uncertain forest areas.
\newblock {\em Journal of the American Statistical Association},
  106(493):31--48.

\bibitem[Finley et~al., 2019]{Finley_nngp}
Finley, A.~O., Datta, A., Cook, B.~D., Morton, D.~C., Andersen, H.~E., and
  Banerjee, S. (2019).
\newblock Efficient algorithms for {Bayesian} nearest neighbor {Gaussian}
  processes.
\newblock {\em Journal of Computational and Graphical Statistics},
  28(2):401--414.

\bibitem[Gaines and Affleck, 2021]{Gaines_Affleck_2021}
Gaines, III, G.~C. and Affleck, D. L.~R. (2021).
\newblock Small area estimation of postfire tree density using continuous
  forest inventory data.
\newblock {\em Frontiers in Forests and Global Change}, Volume 4 - 2021.

\bibitem[Han et~al., 2021]{fire_severity_improvements}
Han, D., Di, X., Yang, G., Sun, L., and Weng, Y. (2021).
\newblock Quantifying fire severity: a brief review and recommendations for
  improvement.
\newblock {\em Ecosystem Health and Sustainability}, 7(1):1973346.

\bibitem[Higuera and Abatzoglou, 2021]{higuera2021record}
Higuera, P.~E. and Abatzoglou, J.~T. (2021).
\newblock Record-setting climate enabled the extraordinary 2020 fire season in
  the {Western United States}.
\newblock {\em Global change biology}, 27(1).

\bibitem[Housman et~al., 2023]{tcc_methods_2023}
Housman, I., Schleeweis, K., Heyer, J., Ruefenacht, B., Bender, S., Megown, K.,
  Goetz, W., and Bogle, S. (2023).
\newblock National land cover database tree canopy cover methods v2021.4.
\newblock GTAC-10268-RPT1. Salt Lake City, UT: U.S. Department of Agriculture,
  Forest Service, Geospatial Technology and Applications Center.

\bibitem[Iglesias et~al., 2022]{Iglesias2022fire}
Iglesias, V., Balch, J.~K., and Travis, W.~R. (2022).
\newblock {U.S.} fires became larger, more frequent, and more widespread in the
  2000s.
\newblock {\em Science Advances}, 8(11):eabc0020.

\bibitem[Jiang et~al., 2023]{Jiang_2023}
Jiang, P., Russell, M.~B., Frelich, L., Babcock, C., and Smith, J.~E. (2023).
\newblock Wildfires correlate with reductions in aboveground tree carbon stocks
  and sequestration capacity on forest land in the {Western United States}.
\newblock {\em Science of The Total Environment}, 893:164832.

\bibitem[Keeley and Syphard, 2021]{2020_california_fires}
Keeley, J. and Syphard, A. (2021).
\newblock Large {California} wildfires: 2020 fires in historical context.
\newblock {\em Fire Ecology}, 17.

\bibitem[Kreider et~al., 2024]{Kreider_2024}
Kreider, M.~R., Higuera, P.~E., Parks, S.~A., Rice, W.~L., White, N., and
  Larson, A.~J. (2024).
\newblock Fire suppression makes wildfires more severe and accentuates impacts
  of climate change and fuel accumulation.
\newblock {\em Nature Communications}, 15(1).

\bibitem[Loehman et~al., 2014]{Loehman_2014}
Loehman, R.~A., Reinhardt, E., and Riley, K.~L. (2014).
\newblock Wildland fire emissions, carbon, and climate: Seeing the forest and
  the trees---a cross-scale assessment of wildfire and carbon dynamics in
  fire-prone, forested ecosystems.
\newblock {\em Forest Ecology and Management}, 317:9--19.

\bibitem[Marlon et~al., 2012]{marlon_wildfire_history}
Marlon, J.~R., Bartlein, P.~J., Gavin, D.~G., Long, C.~J., Anderson, R.~S.,
  Briles, C.~E., Brown, K.~J., Colombaroli, D., Hallett, D.~J., Power, M.~J.,
  Scharf, E.~A., and Walsh, M.~K. (2012).
\newblock Long-term perspective on wildfires in the western {USA}.
\newblock {\em Proceedings of the National Academy of Sciences},
  109(9):535--543.

\bibitem[May and Finley, 2025]{may2025}
May, P.~B. and Finley, A.~O. (2025).
\newblock Spatial–temporal prediction of forest attributes using latent
  {Gaussian} models and inventory data.
\newblock {\em Spatial Statistics}, 69:100917.

\bibitem[Miller et~al., 2009]{Miller_2009}
Miller, J.~D., Knapp, E.~E., Key, C.~H., Skinner, C.~N., Isbell, C.~J., Creasy,
  R.~M., and Sherlock, J.~W. (2009).
\newblock Calibration and validation of the relative differenced {Normalized
  Burn Ratio} ({RdNBR}) to three measures of fire severity in the {Sierra
  Nevada} and {Klamath Mountains}, {California}, {USA}.
\newblock {\em Remote Sensing of Environment}, 113(3):645–656.

\bibitem[Moisen et~al., 2004]{moisen_blackard_finco}
Moisen, G.~G., Blackard, J.~A., and Finco, M. (2004).
\newblock Small area estimation in forests affected by wildfire in the
  {Interior West}.
\newblock {\em Greer, Jerry Dean, ed. Remote sensing for field users;
  proceedings of the tenth Forest Service remote sensing applications
  conference}.

\bibitem[{NASA Jet Propulsion Laboratory}, 2019]{nisar_handbook_2019}
{NASA Jet Propulsion Laboratory} (2019).
\newblock {\em {NASA-ISRO SAR (NISAR) Mission Science Users' Handbook}}.
\newblock National Aeronautics and Space Administration, Pasadena, CA.

\bibitem[Pfeffermann et~al., 2008]{pfeffermann2008}
Pfeffermann, D., Terryn, B., and Moura, F. (2008).
\newblock Small area estimation under a two-part random effects model with
  application to estimation of literacy in developing countries.
\newblock {\em Survey Methodology}, 34.

\bibitem[Polson et~al., 2013]{Polson2013}
Polson, N.~G., Scott, J.~G., and Windle, J. (2013).
\newblock {Bayesian} inference for logistic models using {Pólya–Gamma}
  latent variables.
\newblock {\em Journal of the American Statistical Association},
  108(504):1339--1349.

\bibitem[Prisley et~al., 2021]{Prisley2021}
Prisley, S., Bradley, J., Clutter, M., Friedman, S., Kempka, D., Rakestraw, J.,
  and Sonne~Hall, E. (2021).
\newblock Needs for small area estimation: Perspectives from the {US} private
  forest sector.
\newblock {\em Frontiers in Forests and Global Change}, Volume 4 - 2021.

\bibitem[Reilly et~al., 2022]{hcascadia_2020_wildfires}
Reilly, M.~J., Zuspan, A., Halofsky, J.~S., Raymond, C., McEvoy, A., Dye,
  A.~W., Donato, D.~C., Kim, J.~B., Potter, B.~E., Walker, N., Davis, R.~J.,
  Dunn, C.~J., Bell, D.~M., Gregory, M.~J., Johnston, J.~D., Harvey, B.~J.,
  Halofsky, J.~E., and Kerns, B.~K. (2022).
\newblock {Cascadia} burning: The historic, but not historically unprecedented,
  2020 wildfires in the {Pacific Northwest}, {USA}.
\newblock {\em Ecosphere}, 13(6):e4070.

\bibitem[Sant{\'i}n et~al., 2015]{Santin_2015}
Sant{\'i}n, C., Doerr, S.~H., Preston, C.~M., and Gonz{\'a}lez-Rodr{\'i}guez,
  G. (2015).
\newblock Pyrogenic organic matter production from wildfires: a missing sink in
  the global carbon cycle.
\newblock {\em Global Change Biology}, 21(4):1621--1633.

\bibitem[Shannon et~al., 2024]{Shannon_2025a}
Shannon, E.~S., Finley, A.~O., Domke, G.~M., May, P.~B., Andersen, H.-E.,
  Gaines~III, G.~C., and Banerjee, S. (2024).
\newblock Toward {Spatio-Temporal} models to support national-scale forest
  carbon monitoring and reporting.
\newblock {\em Environmental Research Letters}, 20(1):014052.

\bibitem[Shannon et~al., 2025a]{Shannon2025_biorxiv}
Shannon, E.~S., Finley, A.~O., and May, P.~B. (2025a).
\newblock Quantifying impacts of natural gas development on forest carbon.
\newblock {\em bioRxiv}.
\newblock 10.1101/2025.06.23.661107.

\bibitem[Shannon et~al., 2025b]{Shannon_2025b}
Shannon, E.~S., Finley, A.~O., May, P.~B., Domke, G.~M., Andersen, H.-E.,
  George, III, G.~C., Arne, N., and Banerjee, S. (2025b).
\newblock Leveraging national forest inventory data to estimate forest carbon
  density status and trends for small areas.
\newblock {\em Forest Ecology and Management}, 596:122999.

\bibitem[Stow et~al., 2006]{Stow2006}
Stow, C.~A., Reckhow, K.~H., and Qian, S.~S. (2006).
\newblock A bayesian approach to retransformation bias in transformed
  regression.
\newblock {\em Ecology}, 87(6):1472--1477.

\bibitem[{U.S. Forest Service}, 2019]{usfs_data_discovery_2025}
{U.S. Forest Service} (2019).
\newblock National {USFS} fire occurrence \& national {USFS} fire perimeter.
\newblock {ArcGIS Online} map service.
\newblock Published February 5, 2019; accessed May 19, 2026.

\bibitem[Wear and Coulston, 2015]{wear2015sink}
Wear, D.~N. and Coulston, J.~W. (2015).
\newblock From sink to source: Regional variation in {US} forest carbon
  futures.
\newblock {\em Scientific reports}, 5(1):16518.

\bibitem[Westfall et~al., 2023]{westfall_2023}
Westfall, J.~A., Coulston, J.~W., Gray, A.~N., Shaw, J.~D., Radtke, P.~J.,
  Walker, D.~M., Weiskittel, A.~R., MacFarlane, D.~W., Affleck, D.~L., Zhao,
  D., Temesgen, H., Poudel, K.~P., Frank, J.~M., Prisley, S.~P., Wang, Y.,
  Meador, A. J.~S., Auty, D., and Domke, G.~M. (2023).
\newblock {\em A national-scale tree volume, biomass, and carbon modeling
  system for the {United States}}.
\newblock Gen. Tech. Rep. WO-104. Washington, DC: U.S. Department of
  Agriculture, Forest Service.

\bibitem[Wiener et~al., 2021]{need_for_small_area_estimation}
Wiener, S.~S., Bush, R., Nathanson, A., Pelz, K., Palmer, M., Alexander, M.~L.,
  Anderson, D., Treasure, E., Baggs, J., and Sheffield, R. (2021).
\newblock {United States Forest Service} use of forest inventory data: Examples
  and needs for small area estimation.
\newblock {\em Frontiers in Forests and Global Change}, Volume 4 - 2021.

\bibitem[Woodbury et~al., 2007]{WOODBURY200714}
Woodbury, P.~B., Smith, J.~E., and Heath, L.~S. (2007).
\newblock Carbon sequestration in the {U.S.} forest sector from 1990 to 2010.
\newblock {\em Forest Ecology and Management}, 241(1):14--27.

\bibitem[Wu et~al., 2023]{low_intensity_fires_mitigate_high_intensity_fires}
Wu, X., Sverdrup, E., Mastrandrea, M.~D., Wara, M.~W., and Wager, S. (2023).
\newblock Low-intensity fires mitigate the risk of high-intensity wildfires in
  {California}'s forests.
\newblock {\em Science Advances}, 9(45):eadi4123.

\end{thebibliography}

\end{document}